\documentclass[runningheads]{gd-llncs}
\usepackage{graphicx}
\usepackage{enumitem}
\newcommand{\cit}[1]{``#1''}

\usepackage{color} 

\usepackage{atbegshi}
\usepackage[absolute]{textpos}
\AtBeginShipout{%
   \begin{textblock*}{3mm}(105mm,268mm)
	\textnormal{\the\numexpr\value{page}}
   \end{textblock*}%
}

\begin{document}


\title{Simbanex: Similarity-based Exploration of IEEE VIS Publications}


%
%
\author{Daniel Witschard\inst{1}\orcidID{0000-0001-6150-0787} \and
Ilir Jusufi\inst{1}\orcidID{0000-0001-6745-4398} \and
Andreas Kerren\inst{1,2}\orcidID{0000-0002-0519-2537}}
%
%
\institute{Linnaeus University, Sweden\\
\email{\{daniel.witschard, ilir.jusufi,  andreas.kerren\}@lnu.se} \\
  \and Linköping University, Sweden.\\
\email{\{andreas.kerren\}@liu.se}}
\maketitle              

\begin{abstract}
Embeddings are powerful tools for transforming complex and unstructured data into numeric formats suitable for computational analysis tasks. In this work, we use multiple embeddings for similarity calculations to be applied in bibliometrics and scientometrics. We build a multivariate network (MVN) from a large set of scientific publications and explore an aspect-driven analysis approach to reveal similarity patterns in the given publication data. By dividing our MVN into separately embeddable aspects, we are able to obtain a flexible vector representation which we use as input to a novel method of similarity-based clustering. Based on these preprocessing steps, we developed a visual analytics application, called Simbanex, that has been designed for the interactive visual exploration of similarity patterns within the underlying publications.
\vspace{-1mm}
\end{abstract}
\vspace{-9mm}







\setcounter{footnote}{0}
\section{Introduction} \label{sec:intro}
Embeddings are numeric vector representations of underlying data, and they are normally produced in such a way that items which are similar in the original data set (according to some domain-specific aspect) are embedded into vectors that lie close to each other in the embedding space, with regard to some chosen distance metric. 
The numeric vector format usually makes the embeddings more suitable than the original data as input for computational analysis tasks such as clustering, classification, and similarity calculations. For instance, it is more straightforward to calculate a distance measure, such as Euclidean or cosine distance, with numeric vectors than it is with other types of complex or unstructured data~\cite{Bengio2013Representation,EMB_CLUSTERING1,EMB_CLUSTERING2,EMB_CLUSTERING3}. 
It is important that the embedding algorithms capture the underlying targeted similarity in the best possible way since poor quality embeddings will elevate the risk for low quality results when used in any further calculations. 
Therefore, in recent years a lot of research effort has been dedicated to finding new and better ways to embed different types of data.

For the specific problem of embedding MVNs, current research has explored several methods to embed both the network structure and the attributed data together \cite{NETWORK_SURVEY}. 
However,  separate embedding technologies for data types that are common building blocks for MVNs exist (e.g., separate embeddings for network structure, word/text, categorical attributes etc.)~\cite{GRAPH_SURVEY,WORD_SURVEY}.  
This opens for an alternative approach where different \textit{aspects} of the underlying MVN are first separately embedded and then combined to form the full embedding representation. 

In this paper, we explore such an \textit{aspect-driven} approach on an attributed article citation network built from a large set of scientific publications. On this base we build an interactive application, called Simbanex, which is a further development of our previously proposed experimental tool Simbatex \cite{SIMBATEX}. Simbanex is intended to be used within the fields of bibliometrics and scientometrics and allows the user to perform interactive similarity-based exploration of the underlying set of scientific documents. To demonstrate the usefulness of this type of similarity-driven exploration, we present two different use cases, where the first focuses on citation link analysis and the second on topic similarity. The main trade-off for the application design is between having a heterogeneous framework of several different technologies, or a homogeneous framework based on the same concept (in our case embeddings). We opt for the latter, since we want to explore how far we can come by mainly using, and combining, already existing and well-proven embedding technologies.\\

\noindent The main contributions of our work are:
\begin{enumerate}[leftmargin=*]
    \item A general methodology for dividing a MVN into separately embeddable aspects and strategies to combine the resulting embeddings.
    \item A prototype visual analytics (VA) tool, called Simbanex, which allows the user to explore the similarity patterns of a large set of scientific documents. 
    \item The presentation of two different use cases which illustrate the strengths of our proposed approach and showcase the usage of our tool.
\end{enumerate}


\noindent The rest of this manuscript is organized as follows. In Section~\ref{sec:relwork}, we discuss the relevant related work. 
In Section~\ref{sec:compute}, we describe how we obtain, combine and use the embeddings. The specific details of our proposed VA tools are discussed in Section~\ref{sec:viz} and followed by two use cases that are outlined in Section~\ref{sec:usecase}. 
Finally, in Section~\ref{sec:discussion}, we present the outcomes and limitations of this work.

\section{Related Work} \label{sec:relwork}
Witschard et al. suggest that using and combining several different embeddings could be more beneficial for MVN analysis than using only a single one~\cite{WJMK21}. Since our contribution follows this strategy, we start this section with an overview of the embedding technologies most relevant for our application. We then continue with some of the key aspects from the fields of bibliometrics and scientometrics, and finally, we have a brief look at network visualization.


\noindent \textbf{Word and text embeddings.}\quad
In general, word embeddings are distributed representations obtained from unsupervised training of a deep learning model on some large corpus of natural language text~\cite{Aggarwal2018Machine,Bengio2003ANeural,Bengio2013Representation,Collobert2011Natural,Turian2010Word}. 
By using a large amount of training data to predict words given a specific context (or vice versa), the model will be able to learn semantic similarities of word pairs. 
The algorithm then projects such similar word pairs to embedding vectors that lie close to each other in the embedding space~\cite{WORD_EMBED,WORD_SURVEY}. 
Arguably, the single most influential word embedding technology is Word2Vec, which was introduced in 2013~\cite{W2V}. 
There are different approaches on how to use word embeddings to obtain embeddings for sentences or paragraph-sized text~\cite{Le2014}, starting from the intuitive (but limited) approach to take the average of the embeddings of each word in the text. 
However, sophisticated approaches are needed to exploit the syntactical structure of sentences, which is crucial to do since the same set of words may be arranged to form sentences with different meanings, and the same word may have different meaning depending on the context~\cite{TEXT_EMB1}. 
To do so, the use of deep learning models is a popular choice~\cite{CNN_TEXT,RNN_TEXT,RECURSIVE_TEXT}.
A current state-of-the-art for text embedding is the Universal Sentence Encoder (USE)~\cite{USE} and the BERT model~\cite{BERT}.

 \noindent \textbf{Graph and network embeddings.}\quad
Embedding calculations are not exclusive to textual data. For instance, they can be applied to various important tasks and applications involving graph and network data~\cite{Newman2010Networks,ChuanShi2017ASurvey}.
Technology for graph embedding, also known as Representation Learning on Graphs~\cite{Hamilton2017Representation}, targets the pure topological structure of the graph and ignores any attributed data. 
The goal is to preserve as much as possible of the structure information, and important tasks are clustering, graph comparison, and graph reconstruction~\cite{GRAPH_SURVEY,Grover2016node2vec}. Furthermore, even dynamic aspects can be taken into account for embedding purposes~\cite{Nguyen2018Dynamic}. 
The field of network embedding~\cite{Zhang2020Network} is closely related to the field of graph embedding. 
The main difference is that in addition to the graph topology some (or all) of the attributed data is also considered, which allows for a more elaborated embedding process. 
Consequently, this type of technology is sometimes referred to as Attribute Enhanced Representation Learning~\cite{NETWORK_SURVEY}.

 \noindent \textbf{Bibliometrics and scientometrics.}\quad The concept of bibliometrics can be described as \textit{``the application of mathematical and statistical methods to books and other media”}, and within the subfield of scientometrics the focus lies on analyzing the quantitative aspects of scientific publications and their use. 
So called \textit{distant reading} (i.e., using representations which convey information from the underlying text without the need for actually reading it) is an important concept that has been introduced to alleviate the inherent limitations of normal reading (which in turn is often referred to as \textit{close reading})~\cite{Jaenicke2015OnClose}. Since close reading is time consuming, and time typically is a limiting factor, there is a high demand for distant reading applications which support the navigation of large document sets and convey relevant aggregated information, but still also allow on-demand access to the underlying text for detailed examination. Natural language processing (NLP) in combination with visualization has proved to be a successful combination for tackling such challenges. 
Belinkov and Glass survey the impressive computational progress that has taken place in the field of NLP since the introduction of neural network models~\cite{belinkov-glass-2019-analysis}.
Kucher and Kerren~\cite{Kucher2015Text} provide a taxonomy for, and an overview of, existing methods for text visualization while Liu et al. present a more recent survey on  text mining and visualization~\cite{LWCDOFEJK18}.
The survey of Federico et al.~\cite{Federico2017ASurvey} focuses on visual approaches for analyzing scientific literature and patents while Liu et al.~\cite{Liu2018Scholarly} target visualization and visual analysis of scholarly data.
Finally, the BioVis Explorer by Kerren et al.~\cite{BIOVIS} provides a way to navigate BioVis publications, and their connections based on their respective visualization techniques.

An important observation that is relevant to our work is that it is not uncommon for corpus exploration to be in part driven by questions like \textit{``Are there any groupings of similar documents within the set?”} or \textit{``Are there documents which are similar to this specific document?”}. Therefore, the ability to exploit similarity relations~\cite{TEXTSIM} can be highly relevant for providing useful insights.

\vspace{1ex} \noindent \textbf{Network Visualization.}\quad \sloppy The challenge of exploring network topology in context of the multivariate data attached has received high focus in the last decades~\cite{W2006,Jusufi2013,Kerren14}. As our work does not focus on primarily visualization approaches, but instead tries to integrate embeddings and visualization to explore similarities within the data, discussion of MVN visualization approaches falls outside the scope of our work. However, we advise our readers to check the more recent literature survey on the topic by Nobre et al.~\cite{STAR_MVN_VIZ}.

Visualization can be used to explore network embeddings, as in Li et al.~\cite{LNKHCYM18}, for instance.  However, when it comes to embeddings and visualization of MVNs, especially in terms of combining embeddings, the literature is rather sparse. We have identified this gap in literature and as presented in this paper, our findings suggest this as an interesting new unexplored scientific direction to pursue.

\section{Data Set and Computations} \label{sec:compute}
\begin{figure*}[tb]
    \centering
    \includegraphics[width=\linewidth]{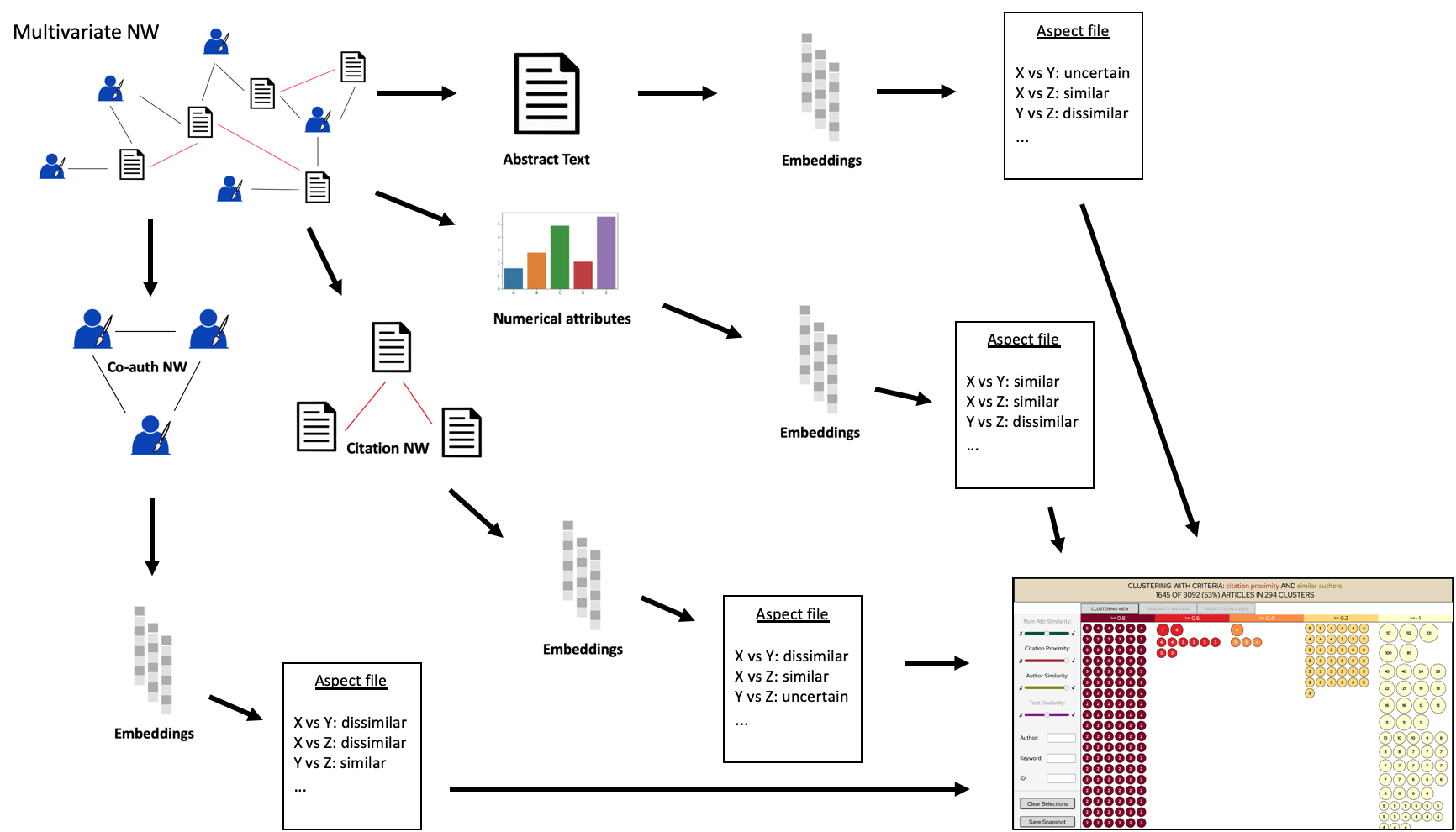}
    \caption{A schematic view of how the aspect-driven approach has been applied to the data set. The underlying MVN is partitioned into four different, node-based \textit{aspects} which are embedded separately. For each aspect, the pairwise similarity classifications (i.e., \textit{similar}, \textit{dissimilar}, or \textit{uncertain}) are calculated. 
    }
    \label{fig:divide_into_aspects}
\end{figure*}

We use the  IEEE VIS data set~\cite{Isenberg2017Vispubdata} which contains information of articles published at the IEEE VIS conferences. 
From this set, we have extracted roughly 3,000 articles published during the period 1990-2018 from which we have built an attributed citation network. We then divide this network into the following 4 node-specific aspects: (1) the article's position in the citation network topology, (2) the abstract text, (3) the co-author information, and (4) two different numerical citation counts. We embed each of these aspects separately and use the embedding vectors to calculate the pairwise cosine similarity scores. These scores are then used to classify each pair into the classes \textit{similar}, \textit{dissimilar}, and \textit{uncertain} with regards to each aspect (see Figure \ref{fig:divide_into_aspects}). 

\noindent One major advantage of this all-embedding strategy is that it gives a straightforward and homogeneous framework for calculating the similarity classifications. 
Furthermore, we note that the methodology is generalizable, beyond the scope of MVNs, since the approach may be used on any complex entity that can be broken down into separately embeddable aspects.

\section{Visualization Approach} \label{sec:viz}
\begin{figure}[htb]
 \centering
 \includegraphics[width=\linewidth]{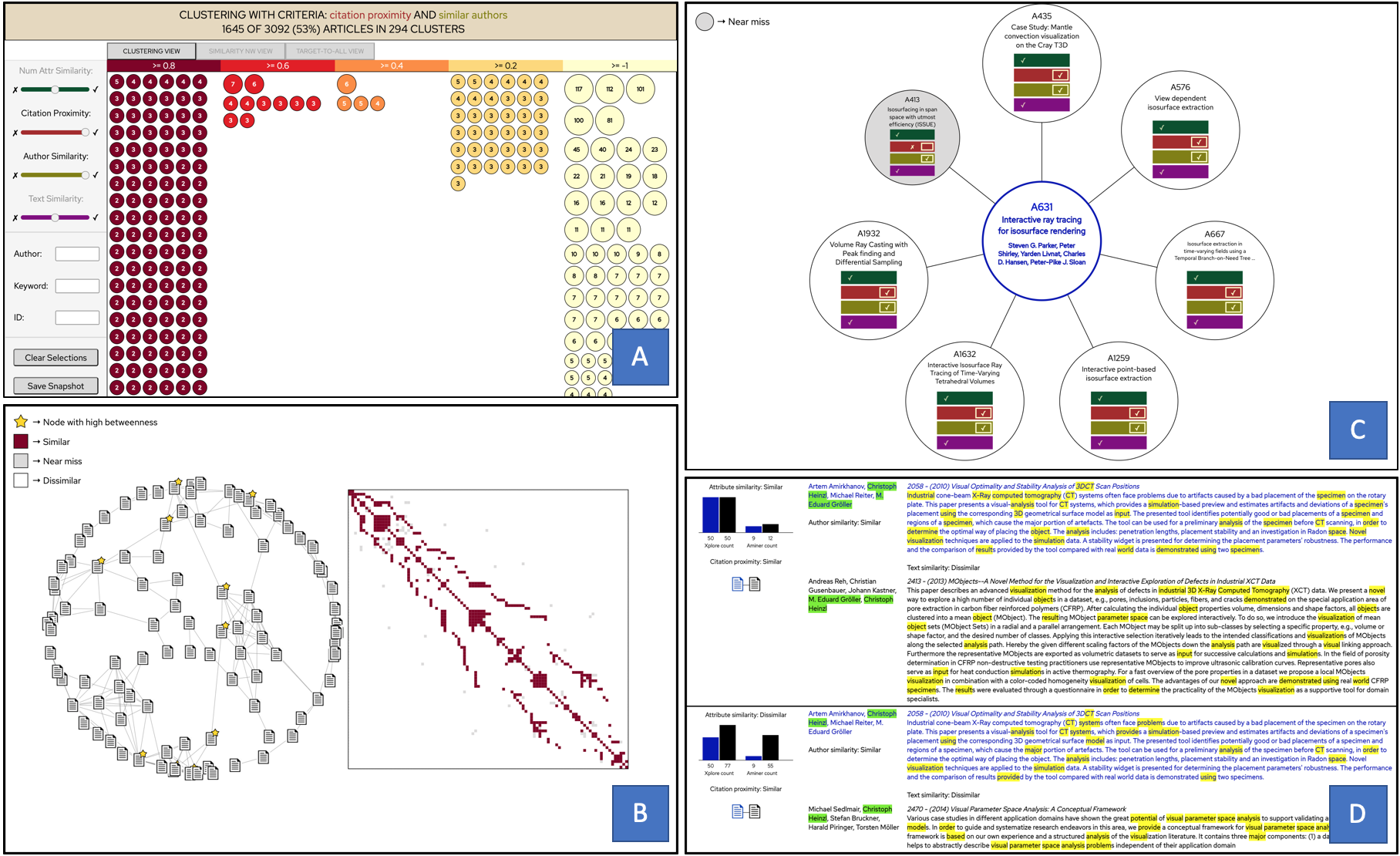}
 \caption{The user interface of Simbanex.
In the \textit{Clustering View} [A], the result of clustering with the current similarity criteria is displayed. In the \textit{Intra Cluster View} [B], the similarity network and the adjacency matrix of a selected cluster can be assessed. The \textit{Target-to-all View} [C] shows an overview of the matches and near misses for a selected article. Finally, the detailed pairwise comparisons can be assessed in the \textit{Similarity Assessment View} [D].
 }
 \label{fig:teaser}
\end{figure}

In this section, we give an overview of the visual design of Simbanex, which is implemented as a web-based tool using D3~\cite{D3}. 
The visualization interface consists of four main views as described in Figure~\ref{fig:teaser}.
Within the application we reuse already well-proven visual metaphors (such as circles for clusters, word-highlighting for text similarity, and node-link diagrams and matrix representations for networks), and it also provides a custom design for the target-to-all comparisons. Furthermore, to facilitate for the user, the visualization continuously provides a textual explanation of the current settings and the current results in the header banner (see Figures~\ref{fig:teaser} [A], \ref{fig:tracking_search}, and \ref{fig:tracking}).

On the conceptual side, we introduce the abstract metaphors of \textit{similarity distribution} and \textit{similarity patterns} (used throughout the remainder of this paper) as mental models for thinking about what happens when a combination of similarity criteria is executed over a set of items. For instance, if we search for pairs with text similarity and citation proximity, we get a different clustering result than if we would search for pairs with text similarity and author dissimilarity (see Section~\ref{subsec:clustering_view}). This could be thought of as the two different criteria specifications having different distributions over the set, in the sense that they each reveal the set of item pairs for which its criteria hold true. Furthermore, we can think of the individual items of the set as being chained together by similarity links that form different patterns depending on the set of active criteria (see  Section~\ref{subsec:sim_nw_view}). 

\subsection{The clustering view}
\label{subsec:clustering_view}
\noindent When the visualization is loaded, the articles are represented as unclustered article icons in the \emph{Clustering View} (see Figure \ref{fig:teaser} [A]). There are four different similarity criteria to use (\textit{Numeric Attribute Similarity}, \textit{Citation Proximity}, \textit{Author Similarity}, and \textit{Text Similarity}), and the user may select yes/no/unactive for each individual criterion. Setting a slider to YES means \textit{``Find all pairs that have been classified as similar for this aspect"}, setting a slider to NO means \textit{``Find all pairs that have been classified as dissimilar for this aspect"}, and setting the slider to the middle, inactive position means \textit{``Do not use this aspect for filtering purposes"}. The user may dynamically select any desired combination to be executed over the data set, and the system will cluster and display all article pairs (if any) that meet the specification. In the terms of our previously introduced terminology, the clustering result is the top-level similarity pattern, and it reveals the similarity distribution of the selected criteria combination over the data set. Clusters are represented as circles where size encodes the number of articles in each cluster, and where spatial position and color both encode the average pairwise similarity score within the cluster. 
Our customized clustering method works as follows: (1) calculate the pairwise similarity with regard to the current criteria, (2) create a similarity matrix where 1/0 in position (N, M) indicates that item N and M are similar/dissimilar, and (3) from the similarity matrix, create clusters so that each article is in the same cluster with all articles which it is similar to. This method does not necessarily give a cluster where all items are similar to all other items within the cluster (which is generally the case for ``normal" clustering). The reason for this is that item X may be similar to item Y (which puts them both into the same cluster) and item Y may be similar to item Z which in turn is dissimilar to X (which still puts Z into the same cluster as Y and X) even. Therefore, all members of a cluster will not always be similar to all others, but they are always connected to each other by at least one path. 

\subsection{The similarity network view}
\label{subsec:sim_nw_view}
\begin{figure}[tb]
    \centering
    \includegraphics[width=\linewidth]{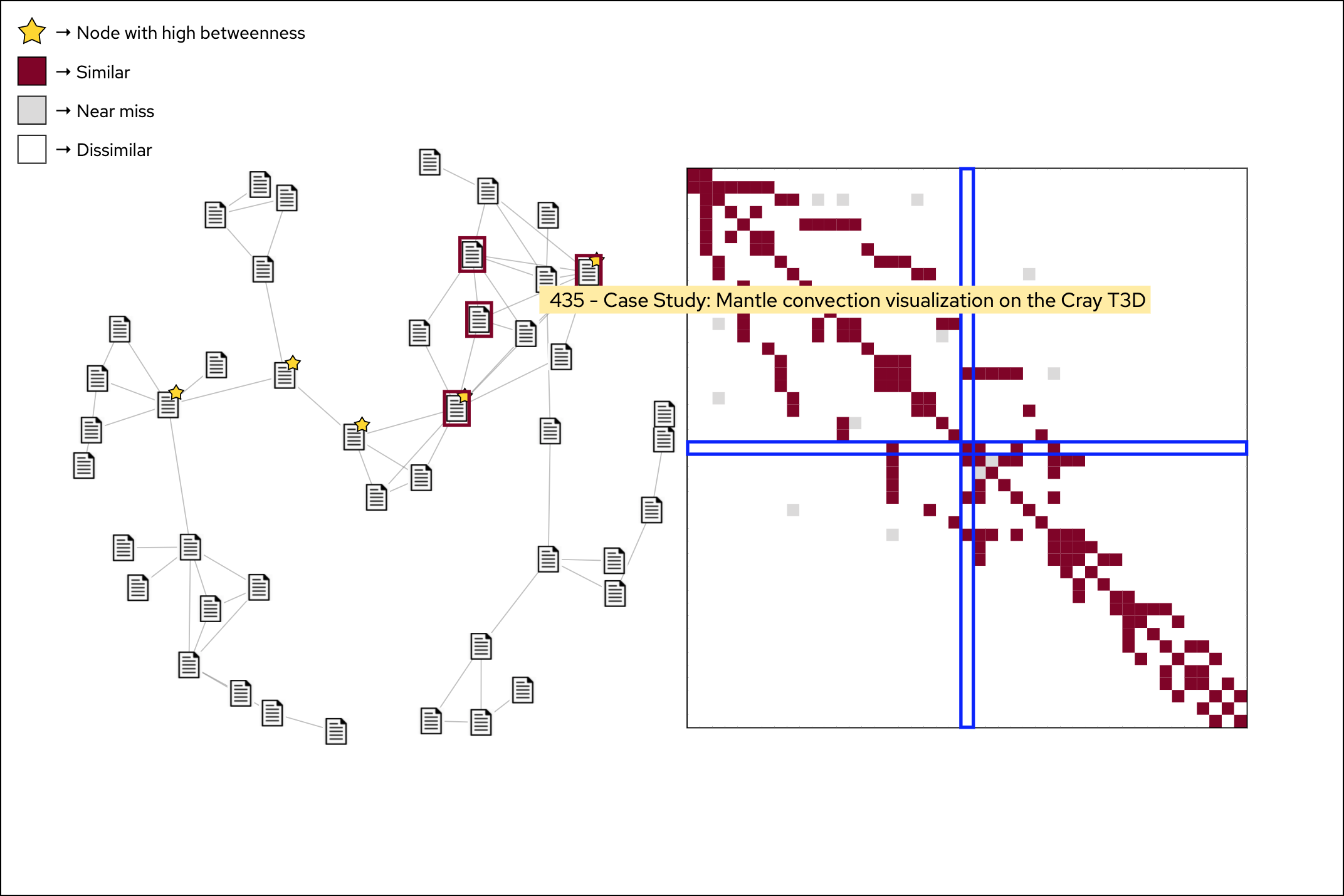}
    \caption{The \emph{Similarity Network View}. Clicking a cluster circle displays the similarity network of the cluster. 
    In this example, the user is hovering the mouse cursor over an article icon to highlight similarity matches and near misses as well as the node's position in the adjacency matrix.
    }
    \label{fig:similarity_network}
\end{figure}

\noindent Clicking a cluster displays the \emph{Similarity Network View} (see Figures~\ref{fig:teaser} [B] and~\ref{fig:similarity_network}) where the similarity links between the articles are displayed both in a node-link diagram and in an adjacency matrix. This view allows for analysis of the similarity pattern of the current cluster which, with regard to our introduced terminology, is an intermediate level pattern. Depending on the selected criteria, different intra-cluster patterns may occur for the same set of items. For example, when using one set of criteria we might get the pattern \textit{``X and Y are similar AND Y and Z are similar BUT X and Z are dissimilar"}, and when using another set of criteria we might get the pattern \textit{``X and Y and Z are all similar"}. Thus, the pattern reveals information on the transitive property of the selected criteria as well as on the overall pairwise homogeneity/heterogeneity within the cluster. This allows for nuanced similarity analysis since items that are not similar when directly compared may still be connected by an indirect ``similarity-path" (and therefore still be similar in some sense). Furthermore, the network pattern/topology can also be used to find items which act as bridges between groups of items with higher inter-connectivity, and Simbanex therefore highlights nodes with a high betweenness centrality with a golden star. 

\subsection{The target-to-all view}
\label{subsec:t2a_view}
\begin{figure}[tb]
    \centering
    \includegraphics[width=\linewidth]{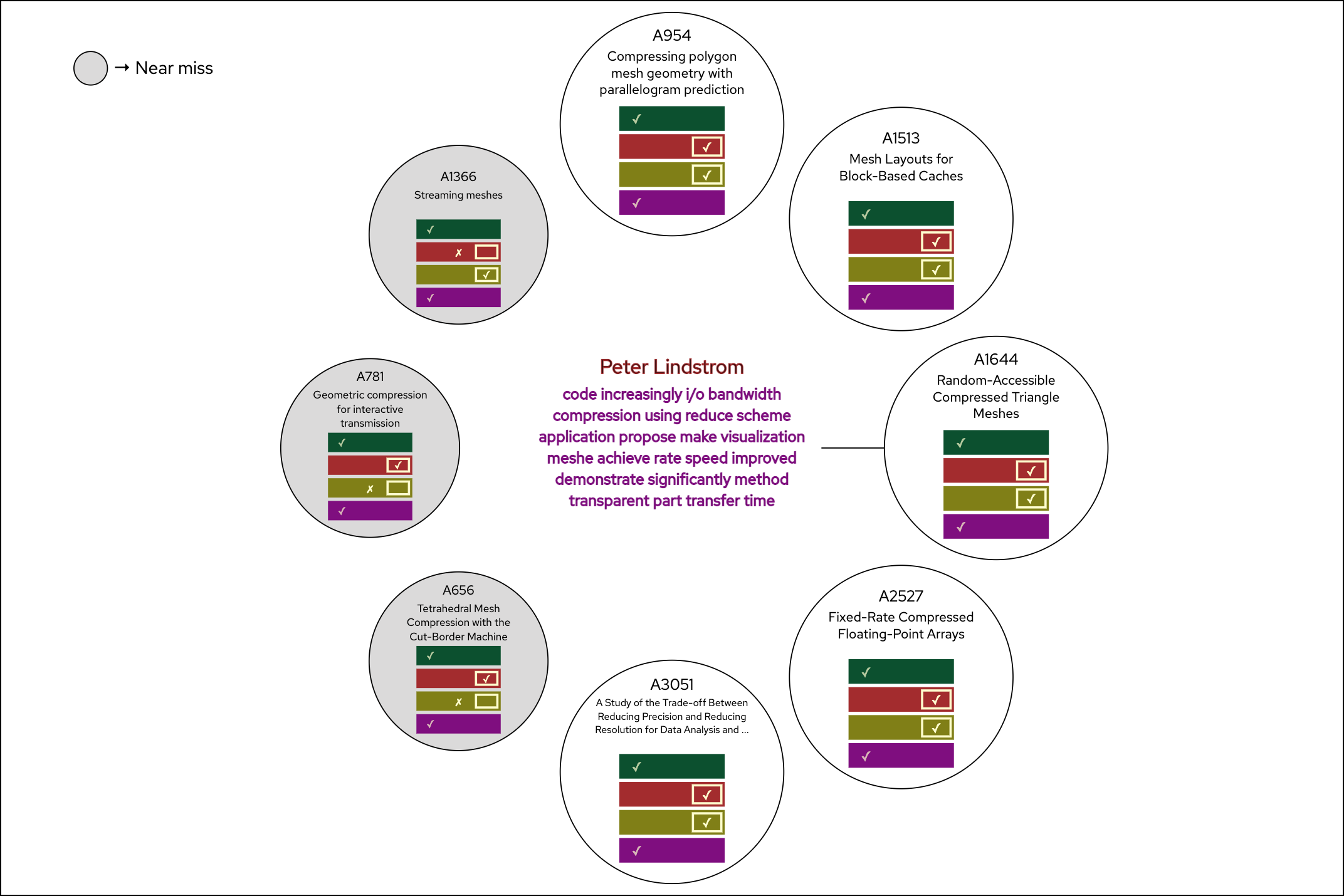}
    \caption{
 The\emph{Target-to-All View} facilitates at-a-glance assessments of matches for a selected target. The colored charts indicate the current setting of activated sliders (the small white frames within the colored areas), and marks indicate whether a full match, or a near miss, was achieved. For non-activated sliders (no white frame present in the colored area), an indication is given for the setting that would result in a match for the corresponding aspect. In this example, there are
5 matches and 3 near misses. The user is hovering the node of the comparison with article A1644 to display co-occurring authors and co-occurring words. 
}
    \label{fig:target_to_all}
\end{figure}

\noindent Clicking an article icon in the \emph{Similarity Network View} displays the \emph{Target-to-All View} (see Figures \ref{fig:teaser} [C] and \ref{fig:target_to_all}) which supports the understanding of how similar the selected target article is to any of the other articles in the data set, given the current criteria settings. The similar items and any near misses are displayed in a radial layout which aims to provide an efficient at-a-glance overview of the pairwise comparison of each aspect. By hovering a comparison node, the user can display any co-occurring authors and any co-occurring words.

\subsection{The similarity assessment view} 
The \emph{Similarity Assessment View} (see Figures \ref{fig:teaser} [D] and \ref{fig:missing_citation}) is displayed just below the \emph{Target-to-All View} and shows the full details of all pairwise comparisons. To facilitate the assessment, the data for the selected target article is color-coded in blue, and co-occurrences of words and authors are highlighted with colored spans. Furthermore, all four system-generated aspect classifications are displayed in such a way that the user can assess them in the direct context of the data.

\subsection{Tracking}
\label{subsec:tracking}
\noindent To support more specific search and analyses, the user may track articles on author name and/or keyword in order to filter the results to show matches only relevant to the current selection. Since matches may still include articles outside of the selection, the tracked articles will be highlighted in green-colored frames throughout the views (see Figure \ref{fig:tracking_search}). In the \emph{Clustering View}, the color coding of the cluster will now indicate the fraction of tracked articles within the cluster (the darker the green the higher the fraction). Spatial position will still encode the average pairwise similarity score within the cluster (see Figure \ref{fig:tracking}). The tracking feature is helpful for answering questions such as if articles that mention certain keywords also show high overall similarity. 

\section{Use Cases} \label{sec:usecase}
Here, we present two use cases to showcase how the methodology could be used as a part of realistic applications in bibliometrics and scientometrics~\cite{Small2006,Yan2009,Federico2017ASurvey}. 
\subsection{Use Case 1 -- Citation link analysis}
\label{subsec:UC1}
\noindent Simbanex makes it possible to interactively explore and get a better understanding of some of the citation patterns within the set of publications. 

\begin{figure}[tb]
    \centering
    \includegraphics[width=\linewidth]{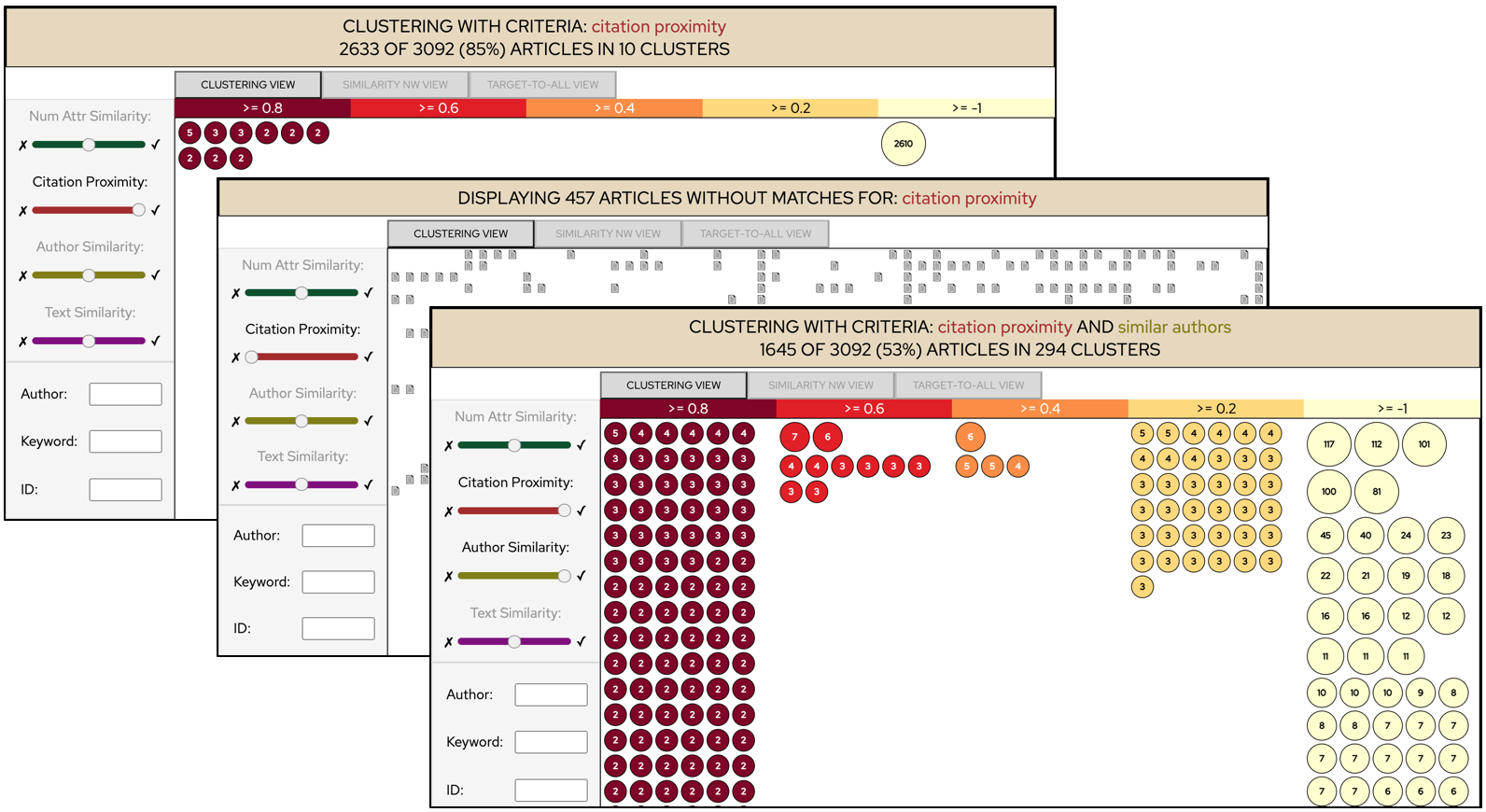}
    \caption{In the first three steps of Use Case 1 (in order from background to foreground), the user explores intra-set citations and self-citations.}
    \label{fig:citation_analysis}
\end{figure}

\vspace{0.75ex} \noindent - Aiming to assess the level of intra-set citations, the user puts the Citation Proximity slider to YES and can see in the banner that there are intra-set citations for roughly 85\% of all publications, see Figure~\ref{fig:citation_analysis} [background].

\vspace{0.75ex} \noindent - Switching the Citation Proximity slider to NO makes it possible to assess the other 15\% of the publications that do not cite publications within the data set and are not cited by any other publication within the data set, see Figure \ref{fig:citation_analysis} [middle]. By browsing the abstracts, the user can discover that an unproportional large amount of these are from early years with regard to the time span of the data set. The user concludes that this is to be expected since these publications have less previous articles to cite within the data, which will substantially lower their probability for having an outgoing citation link. Interestingly enough, very few publications from later years of the time span are found within the subset although the reverse effect (i.e., a lower probability for incoming citation links), which would be expected for these articles. The user therefore concludes that citing within the data set is a trend that has grown stronger in recent years.

\vspace{0.75ex} \noindent - To assess the level of self-citation, the user now sets both the Citation Proximity and the Author Similarity sliders to YES and concludes that the self-citation amounts to about 53\%, see Figure~\ref{fig:citation_analysis} [foreground].

\vspace{0.75ex} \noindent - Setting the Citation Proximity slider to NO and the Text Similarity slider to YES allows the user to explore whether there are any ``missing" citation links between similar publications. This reveals that there are 11 article pairs with high text similarity and no citation link. Eight of these pairs have high pairwise author similarity and three pairs have low pairwise author similarity.

\begin{figure}[h]
    \centering
    \includegraphics[width=\linewidth]{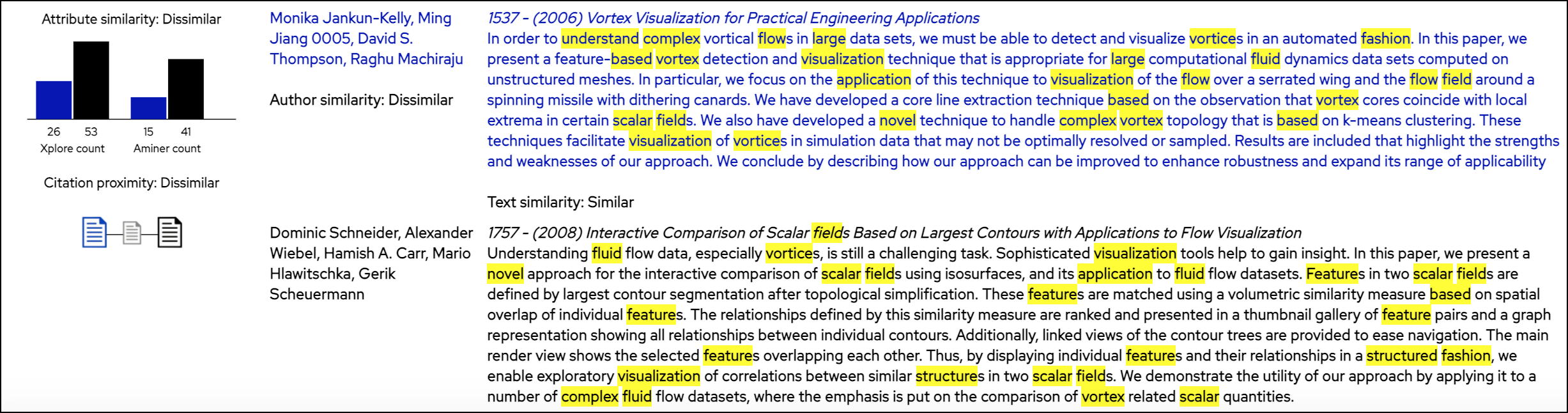}
    \caption{One of the suggestions for \cit{missing citations} from Use Case 1. As can be seen, there already is an indirect (1--hop) link between the two publications, and maybe it could have been relevant to have a direct citation link instead.}
    \label{fig:missing_citation}
\end{figure}

\vspace{0.75ex} \noindent - The user clicks on a cluster of the three with no author similarity to display the similarity network, which in this case is trivial. He then clicks on an article node to assess the detailed pairwise similarity to verify if the match qualifies as a possible citation that should have been made see Figure \ref{fig:missing_citation}.

\begin{figure}[h]
    \centering
    \includegraphics[width=0.80\linewidth]{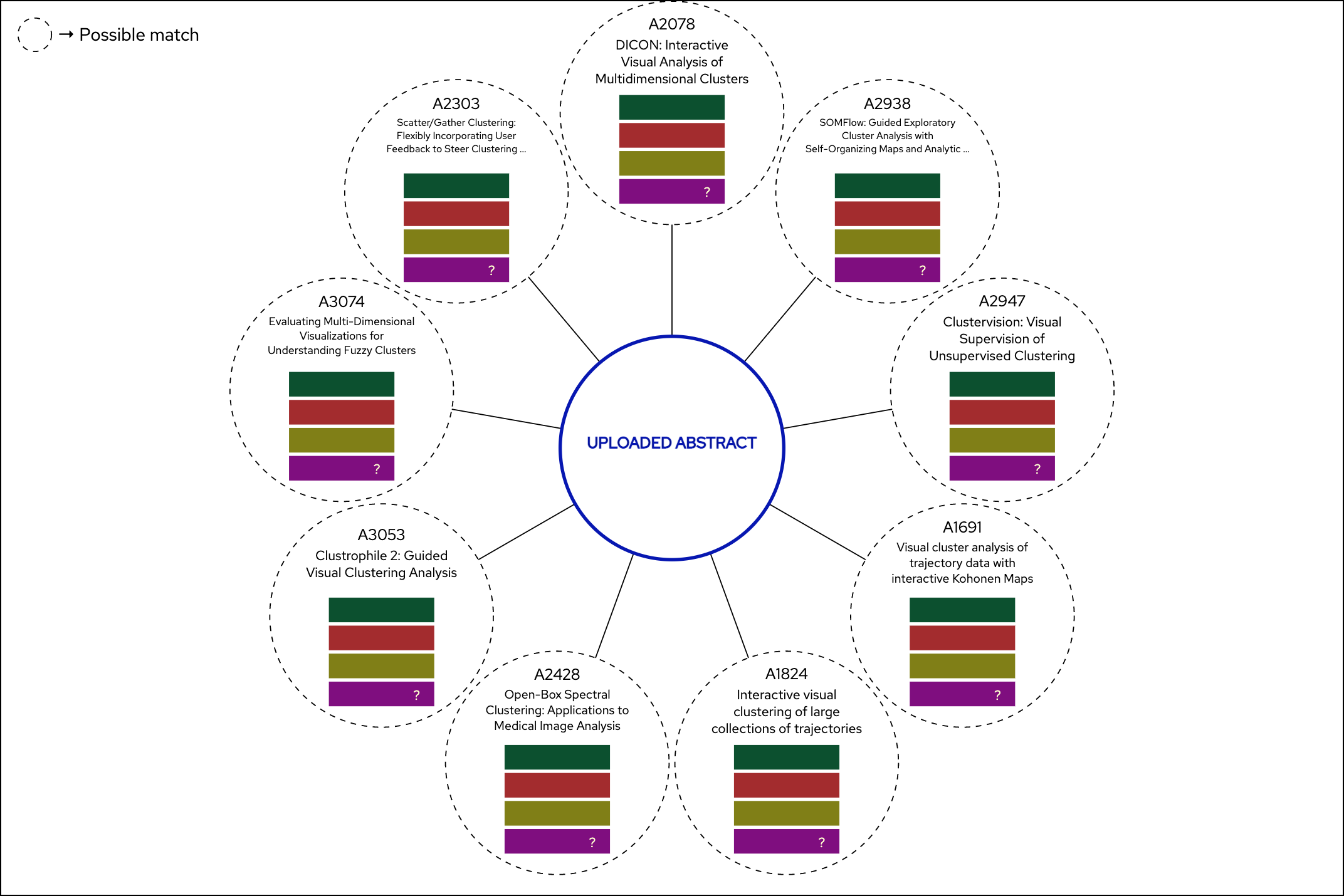}
    \caption{When an abstract is uploaded, the system displays possible matches (if any) based on semantic similarity. The user can then assess each suggestion individually to verify if it makes for relevant citations or not.}
    \label{fig:upload}
\end{figure}

\vspace{0.75ex} \noindent - The user plans to make a submission to an upcoming conference and wants to make sure to cite any previous articles with similar content. He therefore embeds his proposed abstract, puts the resulting files into a specified directory and selects the \textit{Upload Abstract} button. This displays articles with high semantic text similarity (if any) so that an assessment can be made of whether they are relevant candidates for outgoing citations or not, see Figure \ref{fig:upload}.

\clearpage

\subsection{Use Case 2 -- Topic similarity}
\label{subsec:UC2}
\begin{figure}[tb]
    \centering
    \includegraphics[width=\linewidth]{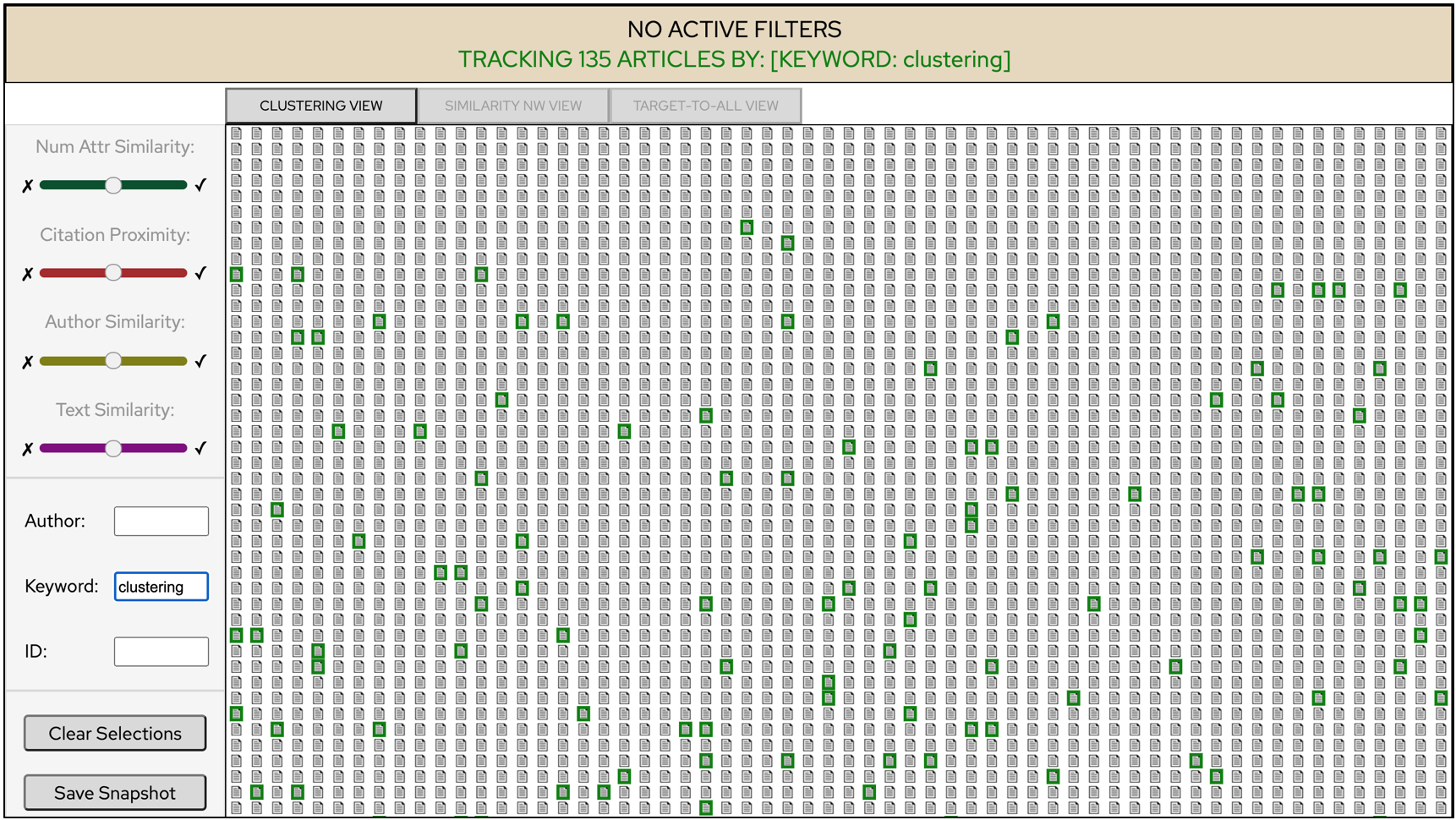}
    \caption{A search for the keyword \emph{clustering} results in a total of 135 publications. When in tracking mode, Simbanex will highlight tracked articles with green frames throughout the views (see Section \ref{subsec:UC2} and Figure \ref{fig:tracking}).
    }
    \label{fig:tracking_search}
\end{figure}

\begin{figure}[tb]
    \centering
    \includegraphics[width=\linewidth]{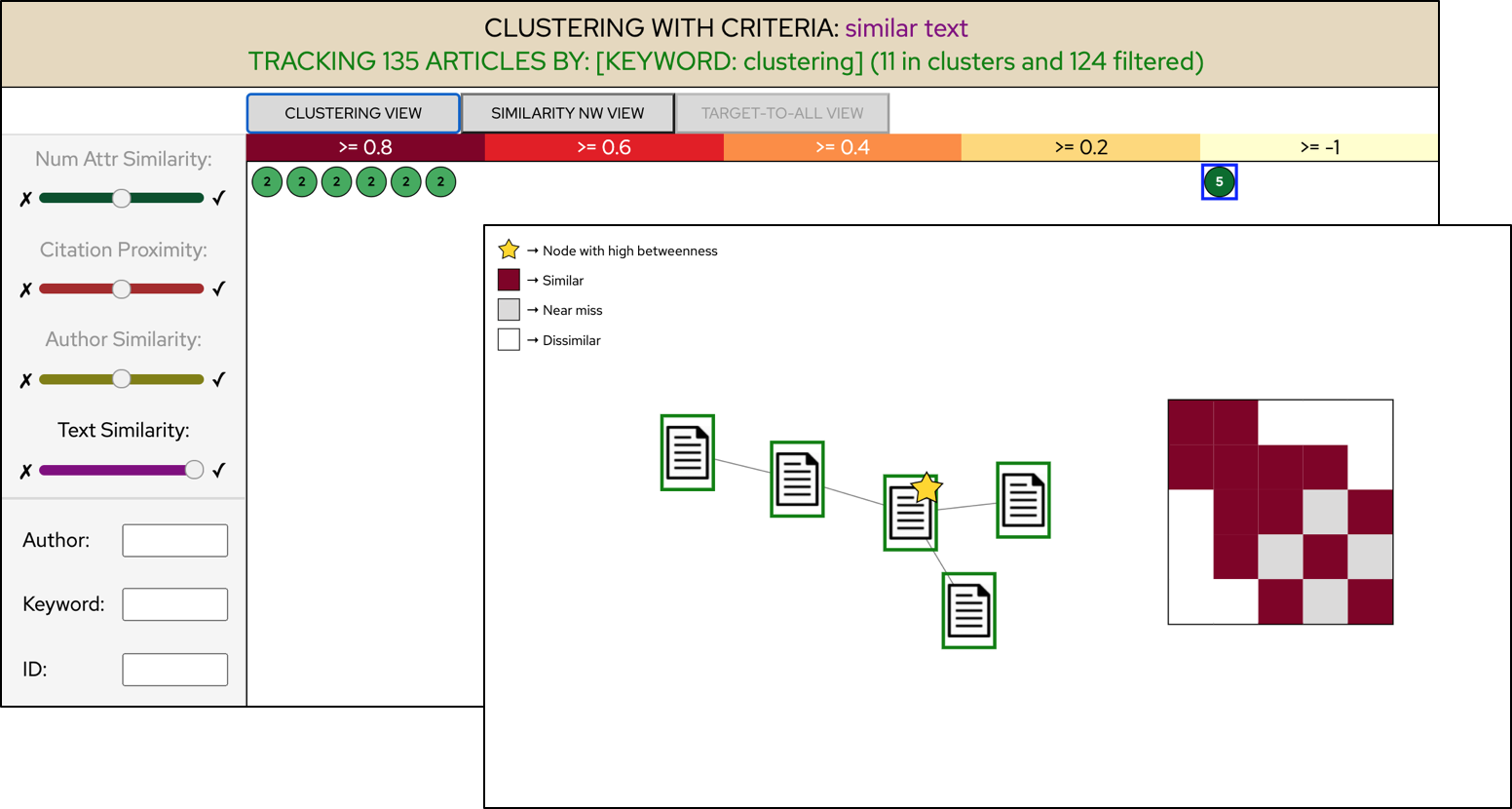}
    \caption{When tracking articles 
    the color intensity of the clusters indicate the fraction of tracked items within the cluster (the darker the green the higher the fraction). Clusters without any tracked articles are filtered. The \emph{Similarity Network View} for the selected cluster reveals that all contained articles are not similar to each other. However, further assessment of the abstract texts shows that they still form a topic cluster on visual cluster analysis (see Section~\ref{subsec:UC2}).}
    \label{fig:tracking}
\end{figure}

\noindent For the second use case, we will use Simbanex to locate topic clusters from selected keywords:

\vspace{1ex} \noindent - In this use case, another user has a special interest in finding out if there are any specific sub groupings within the set of publications for which clustering is an important topic. She therefore enters the keyword \cit{clustering} into the keyword search field and gets a result of a total of 135 articles (which are now highlighted with green frames), see Figure \ref{fig:tracking_search}. 
\vspace{1ex} \noindent - Since it is still not an easy task to assess whether these 135 articles form smaller topic clusters or not (within the larger scope of clustering), the user sets the Text Similarity slider to YES. The system clusters the publications and this time also filters the result so that only clusters containing at least one tracked item remains visible, see Figure \ref{fig:tracking} [background]. 
\vspace{1ex} \noindent - From the intensity of the green color of the clusters, the user concludes that there is one cluster containing a high fraction of tracked articles. However, the average pairwise similarity score indicates lower intra-cluster similarity, which lowers the probability for a sub-topic cluster.
\vspace{1ex} \noindent - Next, the user clicks the cluster to display the similarity network and notes that one article acts like a bridge between 3 of the others, which is a pattern that still makes a topic cluster possible, see Figure \ref{fig:tracking} [foreground].
\vspace{1ex} \noindent - By clicking the article nodes of the similarity network and assessing the abstract texts, the user can conclude that the articles indeed form a cluster on the sub topic of ``visual cluster analysis". Furthermore, for some of the articles there are relevant near misses, so the article set can be expanded further.
\vspace{1ex} \noindent - By noting that the discovered subset of articles would not have been easily selected only by a combination of keywords, the user concludes that there are cases when a similarity-based approach can be used for topic detection.

\section{Discussion and Conclusions} \label{sec:discussion}
In this paper\footnote{This work was partially supported through the ELLIIT environment for strategic research in Sweden.}, we have outlined a methodology for an aspect-driven all-embedding strategy targeting a complex entity such as an MVN. 
The main advantages of the approach is that it provides a homogeneous framework for the similarity calculations and that already existing state-of-the-art embedding algorithms can be used. 
Our methodology generalizes beyond the scope of MVNs, since it is applicable to any type of complex data type that may be broken down into separately embeddable aspects. 
As a proof-of-concept, we have applied the methodology to an attributed citation network using fundamentally different aspects such as network topology, co-author information, text, and numerical data. 

We have built a visual analytics application, called Simbanex, which is designed to be highly responsive and allows the user to explore similarity patterns on different levels of detail. 
With its intended use within the fields of bibliometrics and scientometrics, we have shown how the similarity-based approach can be used to obtain a better understanding of important aspects of the underlying set of scientific publications. This is due to the fact that activities like \textit{\cit{Find other items that are similar/dissimilar to this one with regard to aspects X, Y, and Z.}} are common and important steps in many data exploration tasks. Therefore, Simbanex' ability to interactively activate, negate or disable the similarity criteria one-by-one is highly desirable, since it allows for maximum flexibility when searching for matches. Furthermore, we introduced the novel concept of similarity networks as a way to visualize the associative similarity patterns and obtain a better understanding of the similarity distribution of the current criteria over the data set. Since similarity is an elusive and almost philosophical concept (that very much lies in the eye of the beholder), it is important to stress the fact that a human-in-the-loop solution is needed to combine the strengths of machine-based computing and human pattern recognition---and this has been a main intention when designing our VA tool.

As we have previously discussed, there are cases when it is not possible to break a complex entity down into separately embeddable aspects, which makes it impossible to use the proposed strategy. The scalability of our solution relies on the characteristics of the chosen embedding algorithms and on the size of the data set. While some algorithms allow for asynchronous case-by-case embedding (e.g., USE which is pre-trained), many others rely on embedding all data at once.
We also note that there is an inherent scaling limitation for using pairwise strategies on large data sets, since the number of pairs grows with the square of the number of items.
Finally, we need to point out that the implementation of Simbanex takes advantage of the fact that similarity is relatively scarce within the data set with regards to the current aspects (i.e., most of the article pairs are dissimilar for most of the aspects). In a scenario where similarity would be more common, we would have less of a filtering effect since the comparisons would yield more matches and larger clusters, which in turn would lead to a less responsive user interface.


\bibliographystyle{splncs04}
\bibliography{SimBaNex} 

\end{document}